\documentclass[conference]{IEEEtran}
\IEEEoverridecommandlockouts
\usepackage[english]{babel}
\usepackage{color,soul}
\usepackage{psfrag}
\usepackage{epsfig}
\usepackage{breakcites}
\usepackage{graphicx}
\usepackage[font=footnotesize]{subcaption}  
\usepackage{bm}
\usepackage{cite}
\usepackage{amsmath,amssymb,amsfonts,amsthm}
\usepackage{balance}
\usepackage{textcomp}
\usepackage{cite}
\usepackage{float,algorithm,algpseudocode}
\usepackage{amsopn}\usepackage{placeins}
\usepackage{tikz}
\usepackage{fancyhdr}
\graphicspath{{./img/}}
\DeclareGraphicsExtensions{.pdf,.jpeg,.png}
\def\BibTeX{{\rm B\kern-.05em{\sc i\kern-.025em b}\kern-.08em
    T\kern-.1667em\lower.7ex\hbox{E}\kern-.125emX}}

\newcommand{\calCN}{{\cal CN}}

\newcommand{\calK}{{\cal K}}

\newcommand{\calM}{{\cal M}}

\newcommand{\calS}{{\cal S}}
\newcommand{\calU}{{\cal U}}

\newcommand{\ds}{\displaystyle}

\newcommand{\ba}{\mathbf{a}}

\newcommand{\bd}{\mathbf{d}}
\newcommand{\bg}{\mathbf{g}}
\newcommand{\bh}{\mathbf{h}}
\newcommand{\bp}{\mathbf{p}}

\newcommand{\bs}{\mathbf{s}}
\newcommand{\bw}{\mathbf{w}}

\newcommand{\by}{\mathbf{y}}
\newcommand{\bz}{\mathbf{z}}

\newcommand{\bzero}{\mathbf{0}}

\newcommand{\bA}{\mathbf{A}}

\newcommand{\bD}{\mathbf{D}}

\newcommand{\bG}{\mathbf{G}}
\newcommand{\bH}{\mathbf{H}}
\newcommand{\bI}{\mathbf{I}}

\newcommand{\bQ}{\mathbf{Q}}
\newcommand{\bR}{\mathbf{R}}

\newcommand{\bV}{\mathbf{V}}

\newcommand{\bbE}{\mathbb{E}}
\newcommand{\bbC}{\mathbb{C}}

\newcommand{\test}{{\underset{H_0}{\overset{H_1}\gtrless}}}
\newcommand{\norm}[1]{\left\lVert#1\right\rVert}

\begin{document}

\title{A Comparison Between Co-Located and Distributed MIMO Deployments in OFDM-ISAC Networks
    \thanks{The work of M. Darabi and S. Buzzi was supported by the European Commission through
the Horizon Europe MSCA Doctoral Network project  ISLANDS (grant agreement no. 101120544). The work of Sergi Liesegang was supported by  the European Commission through
the Horizon Europe MSCA Postdoctoral Fellowships project DIRACFEC (Grant No. 101108043).}
}

\author{\IEEEauthorblockN{Maryam~Darabi$^{1,2}$, Sergi~Liesegang$^{2}$, Emanuele~Grossi$^{1,2}$, and Stefano~Buzzi$^{1,2}$}
\IEEEauthorblockA{$^1$\textit{Consorzio Nazionale Interuniversitario per le Telecomunicazioni, 43124 Parma (PR) -- Italia} \\
$^2$\textit{DIEI, Università degli Studi di Cassino e del Lazio Meridionale, 03043 Cassino (FR) -- Italia} }
E-mails: \{maryam.darabi, sergi.liesegang, e.grossi, buzzi\}@unicas.it}

\maketitle
\thispagestyle{fancy} 
\begin{abstract}
This paper investigates network‑level integrated sensing and communication (ISAC) under two fundamentally different topology configurations: cell‑free massive MIMO (CF‑mMIMO) and multi‑cell massive MIMO (MC‑mMIMO). A unified OFDM‑based waveform is adopted for both architectures as the key enabler for ISAC functionalities. The CF system exploits distributed access points (APs) and a scalable user-target‑centric operation, whereas the MC system relies on co‑located transmit–receive arrays with conventional cell‑centric deployment. For both architectures, we derive a GLRT‑based sensing detector and the corresponding sensing SNR expressions. We then examine a series of case studies investigating how the number of OFDM subcarriers, the transceiver allocation strategy, and the antenna/node distribution across the network affect the sensing performance. The results consistently demonstrate that CF‑mMIMO provides more robust and higher sensing performance across most tested scenarios, particularly when transmit resources or antenna elements are spatially distributed. These findings highlight the inherent advantages of CF deployments for next‑generation ISAC networks.
\end{abstract}


\section{Introduction} \label{sec:1}
Integrated sensing and communication (ISAC), as a key enabler in the evolution of both next-generation wireless networks and modern radar systems, has attracted significant research attention \cite{liu2023seventy, meng2024cooperative}. Advances in multiple-input-multiple-output (MIMO) technologies aim at improving spectral efficiency while enabling accurate wireless sensing, thereby supporting the ambitious objectives of ISAC for 6G \cite{li2025sparse}. However, as the focus shifts from link-level to network-level ISAC \cite{meng2025network}, the geographical deployment of network nodes emerges as a critical design challenge:
Distributed MIMO architectures provide rich spatial diversity, enable cooperative processing, and ensure more uniform service across the coverage area \cite{guo2025integrated}. Among these, cell-free massive MIMO (CF-mMIMO) has recently emerged as one of the most promising paradigms \cite{buzzi2026user}.
A CF-mMIMO system consists of a large number of lightweight access points (APs) distributed over a given area and jointly serving user equipments (UEs). In this context, ISAC can fully exploit the intrinsic features of CF-mMIMO, which naturally supports multi-static sensing and removes the need for full-duplex operation \cite{chu2023integrated}. This results in increased sensing degrees of freedom (DoF), enhanced spatial resolution, and improved communication coverage and quality of service (QoS) \cite{femenias2025scalable}.
While initial works on ISAC assumed a simplified single-frequency channel model, more recently the realistic scenario that orthogonal frequency division multiplexing (OFDM) is used has been considered. OFDM is indeed the dominant waveform candidate for 6G systems, and its robustness against frequency-selective fading and its favorable range–Doppler characteristics make it particularly suitable for ISAC applications \cite{wang2025cooperative}.

In this work, we consider two possible network deployments for ISAC: (a) a conventional multi-cell massive MIMO (MC-mMIMO) system \cite{buzzi2019using}, where transmit and receive antennas are co-located and operation is inherently cell-centric, with rigid cell boundaries; and (b) a scalable, user-centric CF-mMIMO architecture \cite{liesegang2025scalable} tailored to network-level ISAC. Assuming, in keeping with the recent trend, the use of OFDM, we provide a direct comparison between these two deployments, highlighting the advantages of CF-mMIMO for joint sensing and communication.

The remainder of the paper is structured as follows. Section~\ref{sec:2} introduces the system model for both deployments. Section~\ref{sec:3} and Section~\ref{sec:4} define the propagation and transmission signals, respectively. Section~\ref{sec:5} details the sensing performance metrics. Section~\ref{sec:6} presents the numerical results, and lastly Section~\ref{sec:7} summarizes the main conclusions.

\section{System Model} \label{sec:2}
In this work, we conduct a network‑level investigation of ISAC, considering two distinct deployment architectures: CF-mMIMO and MC-mMIMO.
The key features and deployment principles of these architectures are elaborated in the subsequent subsections. Within our unified ISAC framework, sensing and communication share the same time–frequency resources, meaning that the transmitted waveforms simultaneously serve both functions without requiring dedicated sensing slots or additional bandwidth. 

\subsection{CF-mMIMO}
Fig.~\ref{fig:1}(a) illustrates the CF-mMIMO architecture wherein a total of $M^{\rm CF}$ geographically distributed APs jointly serve  $K$ single‑antenna UEs for communication while simultaneously probing the environment to detect a single target. Each AP in the CF-mMIMO system is equipped with $N_a^{\rm CF}$-element antenna array, arranged either as a uniform linear array (ULA) or a uniform planar array (UPA),  depending on deployment constraints.  All APs are connected to a central processing unit (CPU) through fully-synchronized high‑capacity fronthaul links, enabling coherent fusion of both communication signals and sensing echoes \cite{liesegang2025scalable}. The CF‑mMIMO architecture supports multistatic ISAC \cite{chu2023integrated} by assigning a subset of APs to transmission (tAPs) and another to reception (rAPs). Let $\cal{M}^{\rm tx}$ and $\cal{M}^{\rm rx}$ denote the sets of tAPs and rAPs within the cooperative region, respectively, with \mbox{$\cal{M}^{\rm tx} \cup \cal{M}^{\rm rx} = \cal{M}^{\rm CF}$} and {$|{\cal{M}}^{\rm CF}|=M^{\rm CF}$}.

To preserve communication-wise scalability, we adopt a user-centric (UC) approach for the CF-mMIMO system \cite{buzzi2017cell}. Instead of relying on a rigid set of serving APs, each UE $k$ is supported by a carefully selected subset of tAPs denoted by $\calM_k^{\rm tx} \subset \calM^{\rm tx}$ with the specific association left unspoken here for the sake of generality. On the other hand, each tAP $m$ serves a subset of UEs indicated by ${\calK}_m^{\rm CF}$.

Furthermore, to ensure sensing-wise scalability,  we adopt a target-centric approach as detailed in \cite{liesegang2025scalable}. Thereby, the entire network footprint is divided into $S$ disjoint sensing zones. Each zone is then further divided into smaller virtual grid cells that span the inspection area. For any virtual cell $\bp_i$ with $i \in S$, we define the subset of tAPs as $\calM_{\bp_i}^{\rm tx} \triangleq \calM_{\bp_i}\cap \calM^{\rm tx}$, which are enabled to simultaneously probe the $S$ virtual cells across the whole area. Denoted by $\calM_{\bp_i}^{\rm rx} \triangleq \calM_{\bp_i}\cap \calM^{\rm rx}$, the subset of rAPs that collect the echoes from the potential target located in $\bp_i$ for the detection task. This structure enables parallel, zone‑wise sensing, where each virtual cell is probed by a dedicated subset of APs. 

\subsection{MC-mMIMO}
\begin{figure}[t]
    \centerline{\includegraphics[scale = 0.15]{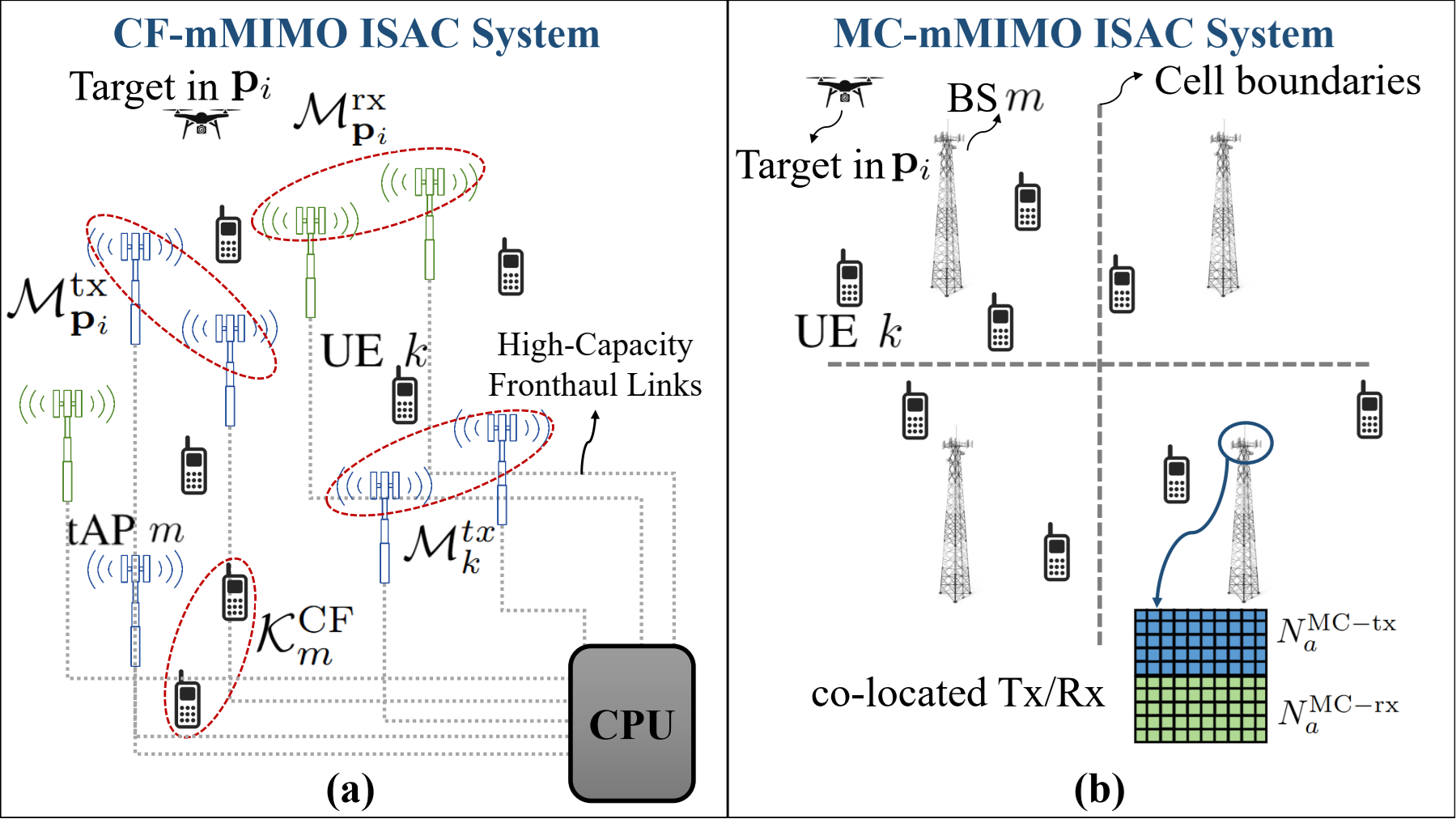}}
    \caption{Deployment of the ISAC system: (a) user‑target‑centric CF‑mMIMO; (b) cell‑centric MC‑mMIMO.}
    \label{fig:1}
\end{figure}

Fig.~\ref{fig:1}(b) depicts the MC-mMIMO system where $M^{\rm MC}$ macro base stations (BSs) are deployed, each located at the center of a cell in the multi‑cell environment. Every BS is equipped with a co‑located transmit–receive antenna array, comprising $N_a^{\rm MC-tx}$ transmit and $N_a^{\rm MC-rx}$ receive elements, arranged either as a ULA or a UPA. This co‑located architecture enables each BS to simultaneously serve its associated $K$ UEs and perform sensing to detect potential targets within its own coverage area. In contrast to the CF‑mMIMO system, where tAPs and rAPs cooperate across the entire network, the MC‑mMIMO architecture follows a strictly cell‑centric operation. Each BS processes its communication and sensing tasks independently, without inter‑BS cooperation.
Accordingly, each BS serves only the UEs located inside its coverage region, denoted by $\calK_m^{\rm MC}$. For the sensing functionality, we define ${\cal{S}}_m^{\rm MC}$ as the set of virtual grid cells assigned to BS $m$ during the detection task. These are the range - azimuth - elevation  radar cells for which the BS $m$ is responsible for detecting a potential target located at position $\bp_i$ with $i \in {\cal{S}}_m^{\rm MC}$. 

\subsection{OFDM Numerology}
At the link level, and independently of the underlying CF‑mMIMO or MC‑mMIMO architecture, all transmitters rely on an OFDM waveform to enable both communication and sensing operations. The occupied bandwidth is $B=N_c \Delta f$ with $N_c$ the number of subcarriers and $ \Delta f$ the subcarrier spacing (SCS). Moreover, the OFDM symbol duration, including the cyclic prefix, is defined as $T_s = 1/\Delta f + T_{\rm CP}$. Each transmission frame contains $N_s$ OFDM symbols, which populate an $N_s \times N_c$ time–frequency grid. Within a coherence block of duration $\tau_c$, the system sequentially carries out uplink (UL) training, UL data transmission, downlink (DL) communication, and DL sensing illumination. Since the OFDM symbol duration is significantly shorter than the coherence interval $T_s \ll \tau_c$, multiple OFDM symbols can be accommodated within a single block, enabling stable channel estimation and high‑resolution sensing.

\section{Propagation Models} \label{sec:3}
Having established the network architectures, we now turn to the corresponding propagation models. The CF‑mMIMO (MC-mMIMO) architecture involves three types of links, including AP–target–AP (BS-target-BS), AP-AP (BS-BS), and UE-AP (BS-UE) channels. The expressions for each channel are provided in the following subsections.
\subsection{CF-mMIMO}
For the target‑associated channel, we assume that only the LoS path exists between the tAP $m$ and rAP $l$. When the target is located at $\bp_{i}$, the corresponding bistatic channel is modeled~as
\begin{equation}
\bH_{m,i,l}(n,n')= \tilde{\alpha}_{m,i,l} \bA_{m,i,l} \rho_i(n)\xi_{m,i,l}(n'),
\label{eq:sensing_Channel}
\end{equation} 
where $n \in \{0, \ldots, N_c - 1\}$, $n' \in \{0, \ldots, N_s - 1\}$ are the OFDM subcarrier and symbol indices, respectively. $\bA_{m,i,l}=\ba_l\left(\varphi_{l,\bp_i},\theta_{l,\bp_i}\right) \ba_{m}^{\rm H}\left(\varphi_{m,\bp_i},\theta_{m,\bp_i}\right)$ is the array response matrix with $\ba_l(\varphi_{l,m},\theta_{l,m})$ the array response vector for the azimuth \mbox{$\varphi_{l,m}$ ($\varphi_{m,l}$)} and elevation \mbox{$\theta_{l,m}$ ($\theta_{m,l}$)} angles of arrival/departure (AoA/AoD) from AP $l$ to AP $m$. The effective gain $\tilde{\alpha}_{m,i,l}=\alpha_{m,i,l} \sqrt{\beta_{m,i,l}}$ accounts for the target reflectivity (RCS) and the large‑scale fading (LSF) along the bistatic link, respectively. The factor $\rho_i(n) = e^{-j2\pi n \Delta f \tau_i}$ represents the phase shift induced by the propagation delay of the target located at $\bp_i$ denoted by $\tau_i$, and $\xi_{m,i,l}(n')=e^{j2\pi n' f_{m,i,l}^d T_s}$ is the Doppler phase shift associated with the target at $\bp_i$, with $f_{m,i,l}^d$ denotes the bistatic Doppler frequency associated to the path $m$-th tAP - target in $\bp_i$ - $l$-th rAP . Assuming the Swerling‑I model, $\alpha_{m,i,l}$ is constant across the OFDM symbols of a coherence block~\cite{behdad2024multi}. 

For the direct propagation path between tAP ($m$) and rAP ($l$)  we adopt a Rician fading model \cite{chu2023integrated}, which described as $\bG_{l,m}(n) = \varkappa_{l,m}\left(\bar{\bG}_{l,m}(n) + \ell_{l,m}(n)\bV_{l,m} \right)$,~where $\varkappa_{l,m} = \sqrt{\beta_{l,m}/(1 + K_{l,m})}$ includes LSF coefficient and Rician K-factor, $\bar{\bG}_{l,m} \in \bbC^{N_a\times N_a}$ represents the NLoS term for which $\bar{\bg}_{l,m} = {\rm vec}(\bar{\bG}_{l,m}) \sim \calCN(\bzero_{N_a^2},\bQ_{l,m})$, $\ell_{l,m}(n) = \sqrt{K_{l,m}}e^{j2\pi \psi_{l,m}(n)}$ with the phase offset $\psi_{l,m}\sim \calU[0,2\pi]$, and $\bV_{l,m} \in \bbC^{N_a \times N_a}$ is the corresponding array response determined by the LoS path.

The uplink channel from UE $k$ to AP $m$, similarly utilizes a Rician model, defined as $\bh_{k,m}(n) =\sqrt{{\beta_{k,m}}/{K_{k,m} + 1}} \left( \sqrt{K_{k,m}} e^{j \psi_{k,m}(n)} \ba_{k,m} + \bh_{k,m}^{\text{SC}}(n) \right)$~where the LOS component consists of the random phase shift and the steering vector from user $k$ to AP $m$, while $\bh_{k,m}^{\text{SC}}(n) \sim \mathcal{CN}(\mathbf{0}_{N_a}, \mathbf{Q}_{N_a})$ models NLoS component. 

\subsection{MC-mMIMO}  
In an analogous way, we now specify the propagation channels for the MC‑mMIMO architecture. The LoS sensing channel associated with the target located at position $\bp_i$, for a co-located transmit-receive array in the $m$-th BS is
\begin{equation}
\bH_{m,i}(n,n')= \tilde{\alpha}_{m,i} \bA_{m,i} \rho_i(n)\xi_{m,i}(n'),
\label{eq:sensing_Channel_MC}
\end{equation} 
where $\tilde{\alpha}_{m,i}$ denotes the effective complex gain, $\rho_i(n)$ captures the delay‑induced phase rotation, and $\xi_{m,i}(n')$ contains the monostatic Doppler shift. The array response matrix is  $\bA_{m,i}=\ba_m\left(\varphi_{m,\bp_i},\theta_{m,\bp_i}\right) \ba_{m}^{\rm H}\left(\varphi_{m,\bp_i},\theta_{m,\bp_i}\right)$ which corresponds to a monostatic configuration where AoA and AoD coincide.

Similar to the CF case, the corresponding BS-BS channel can also be modeled with Rician fading, where $\bG_{l,m}(n) \in \mathbb{C}^{N_a^{\rm MC-rx} \times N_a^{\rm MC-tx}}$ is the channel between BS $m$ and BS $l$ and can be expressed similar to the CF's expression~above. 

To avoid redundancy and due to space constraints, we refrain from deriving the MC communication channels, as detailed in \cite{buzzi2019using}. Here, to keep the notation consistent with the CF case, we denote the uplink channel between BS $m$ and UE $k$ as $\bh_{k,m}(n) \in  \mathbb{C}^{N_a^{\rm MC-tx}}$.

\section{Signal Transmission} \label{sec:4}
Regardless of the network architecture, we consider two unit‑norm OFDM data symbols, $x_{k,m}(n,n')$ and $x_{0,m}(n,n')$, transmitted from the $m$-th tAP/BS for communication with user $k$ and sensing task, respectively. 
For communication, the precoding vector is defined as $\bw_{k,m} = \hat{\bh}_{k,m}^*/\bigl(\mathbb{E}[\|\hat{\bh}_{k,m}\|^2]\bigr)^{1/2}$ where $\hat{\bh}_{k,m}$ is the the minimum mean square error (MMSE) channel estimate between UE $k$ and tAP/BS $m$ obtained as described in \cite[Subsection III-C]{liesegang2025scalable}. For the sensing part, assuming that inspection positions are known, the beamforming vector is simply defined as the steering vector $\bw_{0,m}(\bp_i) = \ba_{i,m}\bigl(\varphi_{m,\bp_i}, \theta_{m,\bp_i}\bigr)$. 

In the CF-mMIMO system, the dual-purpose DL signal transmitted from tAP $m$ is
\begin{equation}
\begin{aligned}
      \mathbf{s}_{m}^{\rm CF}(n,n')&=\sum\nolimits_{k \in {\cal{K}}_{m}^{\rm CF}}  \sqrt{\mu_{k,{m}}} \bw_{k,m} x_{k,{m}}(n,n') \\  &\quad
      + \sum\nolimits_{i \in {\cal S}_{m}^{\rm CF}} \sqrt{\eta_{i,{m}}} \bw_{0,{m}}(\bp_{i}) x_{0,{m}}(n,n').   
\end{aligned}
\label{eq:s_m}
\end{equation}
where $\mu_{k,m}$ and $\eta_{i,m}$ denote the power allocated by tAP $m$ to user $k$ and to the sensing task associated with radar cell $\bp_{i}$, respectively. These coefficients must meet the power constraint $\sum_{k \in {\calK}_{m}^{\rm CF}}\mu_{k,m} + \sum_{i \in {\calS}_{m}^{\rm CF}} \eta_{i,m} \leq P_{m}^{\rm CF}$, where $P_{m}^{\rm CF}$ is the transmit power budget of tAP $m$ \cite{liesegang2025scalable}.

In the cell-centric MC-mMIMO system, the dual-purpose DL signal transmitted from BS $m$ is
\begin{equation}
\begin{aligned}
      \mathbf{s}_{m}^{\rm MC}(n,n')&= \sum\nolimits_{k \in {\cal{K}}_{m}^{\rm MC}} \sqrt{\mu_{k,{m}}} \bw_{k,m} x_{k,{m}}(n,n') \\  &\quad
      + \sum\nolimits_{i \in {\cal S}_{m}^{\rm MC}}  \sqrt{\eta_{i,{m}}} \bw_{0,{m}}(\bp_{i}) x_{0,{m}}(n,n').   
\end{aligned}
\label{eq:s_m_MC}
\end{equation}
In \eqref{eq:s_m_MC}, the sets ${\cal{K}}_{m}^{\rm MC}$ and $ {\cal S}_{m}^{\rm MC}$ denote, respectively, the UEs and the sensing radar cells that are exclusively served by BS $m$. Unlike the user‑target-centric CF architecture, these sets are cell‑centric and disjoint across BSs. The same power constraint applies for MC system as $\sum_{k \in \calK_{m}^{\rm MC}}\mu_{k,m} + \sum_{i \in \calS_{m}^{\rm MC}} \eta_{i,m} \leq P_{m}^{\rm MC}$, where $P_{m}^{\rm MC}$ is the transmit power budget of BS $m$.

\section{Sensing SNR} \label{sec:5}
This section derives the sensing SNR expression, as the key metric used to evaluate both ISAC architectures in the following subsections.
In this work, we focus on sensing~processing; the communication part, including the MC vs. CF comparison, is extensively addressed in prior studies \cite{chu2023integrated, liesegang2025scalable, buzzi2024scalability}.

Moreover, in the following radar processing steps, we assume that the maximum bistatic (or monostatic) Doppler shift satisfies $f^d_{\rm max} \ll \Delta f$ and is known to the system \cite{buzzi2019using}, thus the radar processing considers the maximum Doppler shift.  

\subsection{CF-mMIMO}
Under the hypothesis that a target occupies the virtual‑cell location $\bp_i$, the observation signal captured at rAP~$l\in \calM_{\bp_i}^{\rm rx} = \calM_{\bp_i} \cap \calM^{\rm rx}$ is denoted by \mbox{$\by_l (n,n') \in \mathbb{C}^{N_a}$} and can be written~as
\begin{equation}
    \begin{aligned}
   \by_l(n,n') \approx \sum\nolimits_{m \in \calM^{\rm tx}}\bH_{m,i,l}\bs_{m}^{\rm CF}(n,n') +  \bz_l(n,n').
    \end{aligned}
    \label{eq:sensing-received-signal}
\end{equation}
where \mbox{$\bz_l(n,n')\sim \calCN(\bzero, \sigma^2_z \bI_{N_a})$} is thermal noise. Given that we assume tight synchronization enforced by the CPU, the AP–AP links and the transmitted waveforms are known and can therefore be suppressed before radar processing \cite{chu2023integrated}. That is why no additional clutter has been explicitly included in writing \eqref{eq:sensing-received-signal}\footnote{However, as shown in \cite{liesegang2025scalable}, imperfect AP‑AP interference cancellation (clutter) induces a significant performance loss. Its explicit modeling and mitigation are thus left for future work.}. This assumption is common in cooperative ISAC architectures and enables a cleaner characterization of the target return.

Following the sensing processing proposed~in~\cite{liesegang2025scalable},~we extend~the~formulation~for~an~OFDM-based~structure. To express the sensing model~compactly,~we~define spatial~sensing response matrix $\bD_{i,l}(n,n')\triangleq \bigl[\bd_{m,i,l}(n,n')\bigr]_{m \in \calM^{\rm tx}} \in \mathbb{C}^{N_a \times |\calM^{\rm tx}|}$,~with $\bd_{m,i,l}(n,n') = \sqrt{\beta_{m,i,l}} \bA_{m,i,l} \rho_i(n) \xi_{m,i,l}(n') \bs_{m}^{\rm CF}(n,n')$, and vectorize~the RCS coefficients across~tAPs~as~$\bm{\alpha}_{i,l}\triangleq \bigl[\alpha_{m,i,l}\bigr]_{m \in \calM^{\rm tx}} \in \mathbb{C}^{|\calM^{\rm tx}|}$. Thus,~\eqref{eq:sensing-received-signal} can be rewritten as 
\begin{equation}
 \by_l(n,n') \approx  \bD_{i,l}(n,n') \bm{\alpha}_{i,l} + \bz_l(n,n').
     \label{eq:sensing-received-signalFinal}
\end{equation}
Next we stack the observation \eqref{eq:sensing-received-signalFinal} over the entire OFDM frame $N_s \times N_c$ and obtain vector and matrices denoted as  $\ddot{\by}_{i,l}$, $\ddot{\bD}_{i,l}$, and $\ddot{\bz}_l$. The radar detection test for the target in position $\mathbf{p}_i$  becomes the following binary hypothesis test:
\begin{equation}
    \left\{\begin{array}{ll}
        H_1: & \ddot{\by}_{i,l} = \ds \ddot{\bD}_{i,l} \bm{\alpha}_{i,l} + \ddot{\bz}_l \\
        H_0: & \ddot{\by}_{i,l} = \ds \ddot{\bz}_l ,
    \end{array} \right.
    \label{eq:hypothesis}
\end{equation}

To address the hypothesis test in \eqref{eq:hypothesis}, we treat the unknown vector $\bm{\alpha}_{i,l}$ as deterministic and therefore employ the generalized likelihood ratio test (GLRT), which yields
\begin{equation}
    \frac{\max_{\bm{\alpha}_{i,l}} f(\ddot{\by}_{i,l} | H_1, \bm{\alpha}_{i,l})}{f(\ddot{\by}_{i,l} | H_0)} \gtrless \delta_i.
    \label{eq:GLRT-doppler}
\end{equation}
The threshold $\delta_i$ is set to guarantee that the false‑alarm probability for the $i$-th detection test does not exceed the specified level. Note that unlike \cite{behdad2024multi}, our formulation uses no prior knowledge about the RCS statistics, thereby making the processing more generic and not tailored to any specific distribution.
Applying the maximum‑likelihood criterion to $\bm{\alpha}_{i,l}$ results in the estimate
\begin{equation}
     \hat{\bm{\alpha}}_{i,l} = \ddot{\bD}_{i,l}^\dagger \ddot{\by}_{i,l},
 \label{eq:ML-alpha}
\end{equation} 
with \mbox{$\ddot{\bD}_{i,l}^\dagger=(\ddot{\bD}_{i,l}^{\rm H}\ddot{\bD}_{i,l})^{-1} \ddot{\bD}_{i,l}^{\rm H}$}. Then, substituting \eqref{eq:ML-alpha} in \eqref{eq:GLRT-doppler}, we will have 
\begin{equation}
    \sum\nolimits_{l \in \calM_{\bp_i}^{\rm rx}} \norm{\ddot{\bD}_{i,l}\ddot{\bD}_{i,l}^\dagger\ddot{\by}_{i,l}}^2 \test \delta'_i,
    \label{eq:Final-GLRT}
\end{equation}  
with $\delta'_i=\sigma_z^{2} \ln{\delta_i}$. In \cite{liesegang2025scalable}, the receive SNR is shown to be closely related to the detection probability, which is why, in our formulation, we use SNR purely as a signal‑quality metric derived from the GLRT statistic. Based on this interpretation, we define the received sensing SNR for the $i$-th sensing region in the CF architecture as
\begin{equation}
    \gamma_{\bp_i}^{\rm CF}
    = \sum_{l \in \calM_{\bp_i}^{\rm rx}}
        {\rm tr} \big(\ddot{\bD}_{i,l}\bR_{i,l}\ddot{\bD}_{i,l}^{\rm H}
        \big)/\bigg(\sigma_z^2
        \sum_{l \in \calM_{\bp_i}^{\rm rx}} r_{i,l}\bigg),
    \label{eq:gamma_detection_single}
\end{equation}
where $r_{i,l}={\rm rank}(\ddot{\bD}_{i,l})$, and $\bR_{i,l}= \bbE[\bm{\alpha}_{i,l}\bm{\alpha}_{i,l}^{\rm H}]$ is the covariance matrix of $\bm{\alpha}_{i,l}$. 

\subsection{MC-mMIMO}
We apply the same radar‑processing framework to the MC‑mMIMO system in an analogous manner. For BS $m$, which employs co‑located transmit and receive arrays, the received observation $\by_m(n,n') \in \mathbb{C}^{{N_a}^{\rm rx}}$ can be written as 

\begin{equation}
    \begin{aligned}
   \by_m(n,n') \approx \bH_{m,i} \bs_{m}^{\rm MC}(n,n') +  \bz_m(n,n').
    \end{aligned}
    \label{eq:sensing-received-signal1-MC}
\end{equation}
where noise is modeled as \mbox{$\bz_m(n,n')\sim \calCN(\bzero, \sigma^2_z \bI_{{N_a}^{\rm rx}})$}. In line with standard monostatic sensing models \cite{he2022joint}, given the cell‑centric operation of the MC‑mMIMO system, and for the fairness with CF case in this work, we do not include inter‑cell interference or clutter in the sensing formulation\footnote{In general, inter‑cell interference can be modeled as additional noise \cite{chen2019multi}; therefore, our work can be readily extended to that scenario.}. Since each BS performs sensing only within its own coverage area and neighboring BSs are geographically separated, their downlink transmissions are assumed to have a negligible impact on the local radar return. This assumption keeps the sensing model compact while remaining consistent with prior ISAC analyses.  
The spatial sensing vector for the MC system then becomes $\bd_{i,m}(n,n')\triangleq [\sqrt{\beta_{m,i}} \bA_{m,i} \rho_i(n) \xi_{m,i,l}(n') \bs_{m}^{\rm MC}(n,n')] \in \mathbb{C}^{N_a^{\rm rx}}$, and the scalar RCS coefficients is ${\alpha}_{i,m}$. Hence, consistent with CF notation, we have 
\begin{equation}
 \by_m(n,n') \approx  \bd_{i,m}(n,n') {\alpha}_{i,m} + \bz_m(n,n'),
     \label{eq:sensing-received-signal2-MC}
\end{equation}
Similar to previous subsection, we stack the observations in \eqref{eq:sensing-received-signal2-MC} over $N_s$ symbols and $N_c$ subcarriers and thus the $i$-th test becomes
\begin{equation}
    \left\{\begin{array}{ll}
        H_1: & \ddot{\by}_{i,m} = \ds \ddot{\bd}_{i,m} {\alpha}_{i,m} + \ddot{\bz}_m \\
        H_0: & \ddot{\by}_{i,m} = \ds \ddot{\bz}_m ,
    \end{array} \right.
    \label{eq:hypothesis_MC}
\end{equation}

Like before, in \eqref{eq:hypothesis_MC}, ${\alpha}_{i,m}$ is assumed to be deterministic and the GLRT becomes 
\begin{equation}
    \frac{\max_{{\alpha}_{i,m}} f(\ddot{\by}_{i,m} | H_1, {\alpha}_{i,m})}{f(\ddot{\by}_{i,m} | H_0)} \gtrless \delta_i.
    \label{eq:GLRT-MC}
\end{equation}
Accordingly, the ML estimate of ${\alpha}_{i,m}$ is computed as  $ \hat{{\alpha}}_{i,m} = \ddot{\bd}_{i,m}^H \ddot{\by}_{i,m}/{\norm{\ddot{\bd}_{i,m}}^2}$. The GLRT for the MC system then becomes 
 
\begin{equation}
{\lvert{\ddot{\bd}_{i,m}^H \ddot{\by}_{i,m}\rvert^2}}/{{\norm{\ddot{\bd}_{i,m}}^2}} \test \delta'_i.
    \label{eq:Final-GLRT-MC}
\end{equation}  
The GLRT in \eqref{eq:Final-GLRT-MC} and corresponding sensing SNR expressions for the MC case coincide with the classical scalar‑amplitude radar detector (e.g., \cite{buzzi2022foundations}), where the test statistic reduces to the energy of the projection of the received vector onto the steering direction. Following the formulation in \cite{buzzi2022foundations}, the corresponding sensing SNR becomes
\begin{equation}
    \gamma_{\bp_i}^{\rm MC}
    ={\sigma_{\alpha_{i,m}}^2{\norm{\ddot{\bd}_{i,m}}^2}}/{\sigma_z^2},
    \label{eq:gamma_detection_single_MC}
\end{equation}
where $\sigma_{\alpha_{i,m}}^2= \bbE[{\alpha}_{i,m}{\alpha}_{i,m}^{*}]$ denotes the variance of the monostatic scattering coefficient.

\section{Numerical Simulations} \label{sec:6}
In our numerical simulations, we consider a $1  \text{km}^2$ service area for both ISAC network architectures. A total of $K=16$ UEs are distributed randomly across the area with their height fixed to $1.65 \text{m}$. The target location is also drawn uniformly at random, with an elevation ranging from $20 \text{m}$ to $100 \text{m}$. The system operates at a carrier frequency $f_c = 3\text{GHz}$, employing a total of $N_c=12$ subcarriers and $N_s=14$ OFDM symbols with SCS $\Delta f = 30 \text{kHz}$. 
A ULA configuration is adopted throughout the entire simulation setup. A uniform power allocation is applied across the target and all UEs, subject to the constraint $\sum_{m \in M^{\rm BS}} P_m^{\rm MC} = \sum_{m \in \calM^{\rm tx}} P_m^{\rm CF}$ ensuring equal total transmit power in the MC and CF architectures. The noise power spectral density is set to $N_0=-174 \text{dBm/Hz}$ and the receiver noise figure is $9 \text{dB}$, resulting in a noise variance $\sigma_z^2 = N_0 B$. The target’s RCS is modeled as ${\alpha}_{i,m,m'} \sim \calCN(0,10 \quad \rm dBsm)$. 
To guarantee a fair comparison between the CF and MC networks, we impose $M^{\rm MC} = N_a^{\rm CF}$, $N_a^{\rm MC-tx}=\lvert {\calM^{\rm tx}}\rvert$, $N_a^{\rm MC-rx}=\lvert{\calM^{\rm rx}}\rvert$, and $P_m^{\rm MC} = M^{\rm CF} P_m^{\rm CF}/M^{\rm MC}$, unless otherwise stated. 

In the CF-mMIMO system, we consider $S=4$ sensing zones and $N_a^{\rm CF}=4$ antennas per APs. The APs are randomly deployed within the service area at a height of $10 \text{m}$, and the specific number of tAP and rAP are reported for each numerical experiment. In the MC-mMIMO system, we have $4$ non-overlapping cells, and each BS is located at the center of each cell with a height of $10 \text{m}$. 
In the following subsections, we present three studies to evalute the performance of the network‑level ISAC frameworks \footnote{ While \cite{liesegang2025scalable, buzzi2024scalability} investigate the communication performance and report the expected robustness, a comprehensive analysis tailored to the considered case studies is left for future work.}.

\subsection{Distributed vs. Co‑Located: Frequency Diversity Impact}
\begin{figure}[t]
    \centerline{\includegraphics[scale = 0.26]{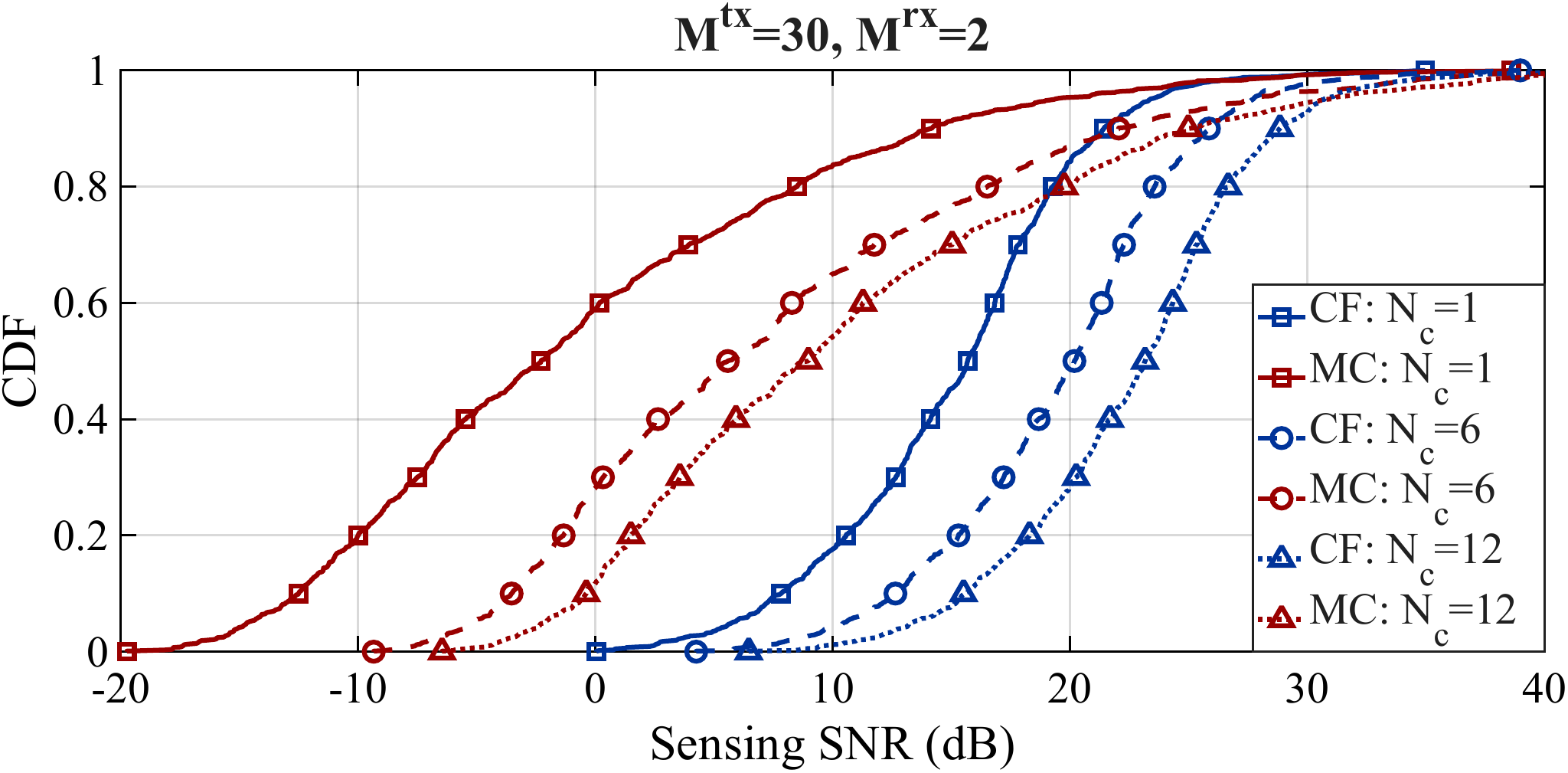}}
    \caption{Sensing SNR CDF  w.r.t $N_c$ in both ISAC network frameworks for transceiver split $\lvert{\calM^{\rm tx}}\rvert =32$ and $\lvert{\calM^{\rm rx}}\rvert =2$.}
    \label{fig:2}
\end{figure}

\begin{figure}[t]
    \centerline{\includegraphics[scale = 0.26]{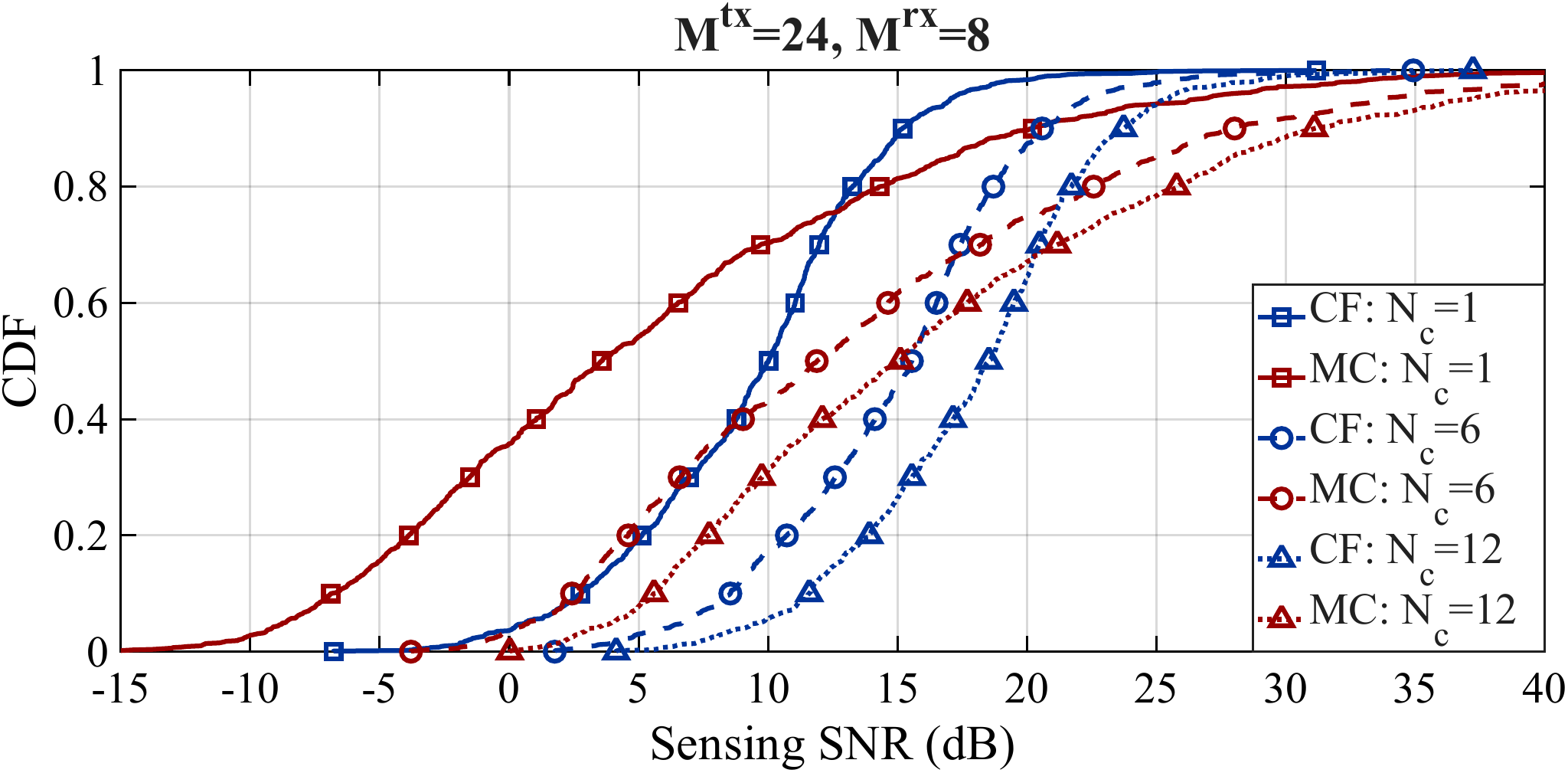}}
    \caption{Sensing SNR CDF w.r.t $N_c$ in both ISAC network frameworks for transceiver split $\lvert{\calM^{\rm tx}}\rvert =24$ and $\lvert{\calM^{\rm rx}}\rvert =8$.}
    \label{fig:3}
\end{figure}

We consider two transceiver‑allocation settings for both architectures and evaluate their sensing performance while varying the number of subcarriers. The two configurations are defined as $N_a^{\rm MC-tx}=\lvert {\calM^{\rm tx}}\rvert = \{24,30\}, N_a^{\rm MC-rx}=\lvert{\calM^{\rm rx}}\rvert= \{8,2\}$ with the total number of APs fixed at $M^{\rm CF}=32$. Fig.~\ref{fig:2} and ~\ref{fig:3} report the received sensing SNR for different numbers of subcarriers ($N_c$) under these two transceiver splits. In the CF‑ISAC architecture, all APs cooperatively illuminate and sense the target, whereas in MC‑ISAC, a single BS equipped with a co‑located antenna array performs both functions. When the transmit‑heavy configuration is used (Fig.~\ref{fig:2}), the CF architecture consistently achieves a higher sensing SNR at the 95‑percentile level for all tested values of $N_c$, despite the BS having a much larger physical aperture. For the more balanced configuration (Fig.~\ref{fig:3}), CF outperforms MC in roughly $80\%$ of the cases when $N_c=1$, and still in about $70\%$ when $N_c=12$. For both architectures, increasing the number of subcarriers improves the sensing SNR, since the additional frequency diversity provides multiple partially independent observations of the target, reducing the impact of frequency‑selective fading and yielding a more stable aggregated return.
Furthermore, in CF-ISAC, the spatial diversity and geometric richness inherent to its distributed topology, wherein each AP contributes an independent bistatic sensing path, and the aggregation of many weak but spatially diverse returns yields a highly stable composite echo. In contrast, the MC‑ISAC system relies on a single co‑located array. Although this array provides strong coherent gain, its performance is more sensitive to unfavorable target–BS geometries, resulting in larger variation.
\subsection{Distributed vs. Co‑Located: Transceiver Allocation}        
\begin{figure}[t]
    \centerline{\includegraphics[scale = 0.27]{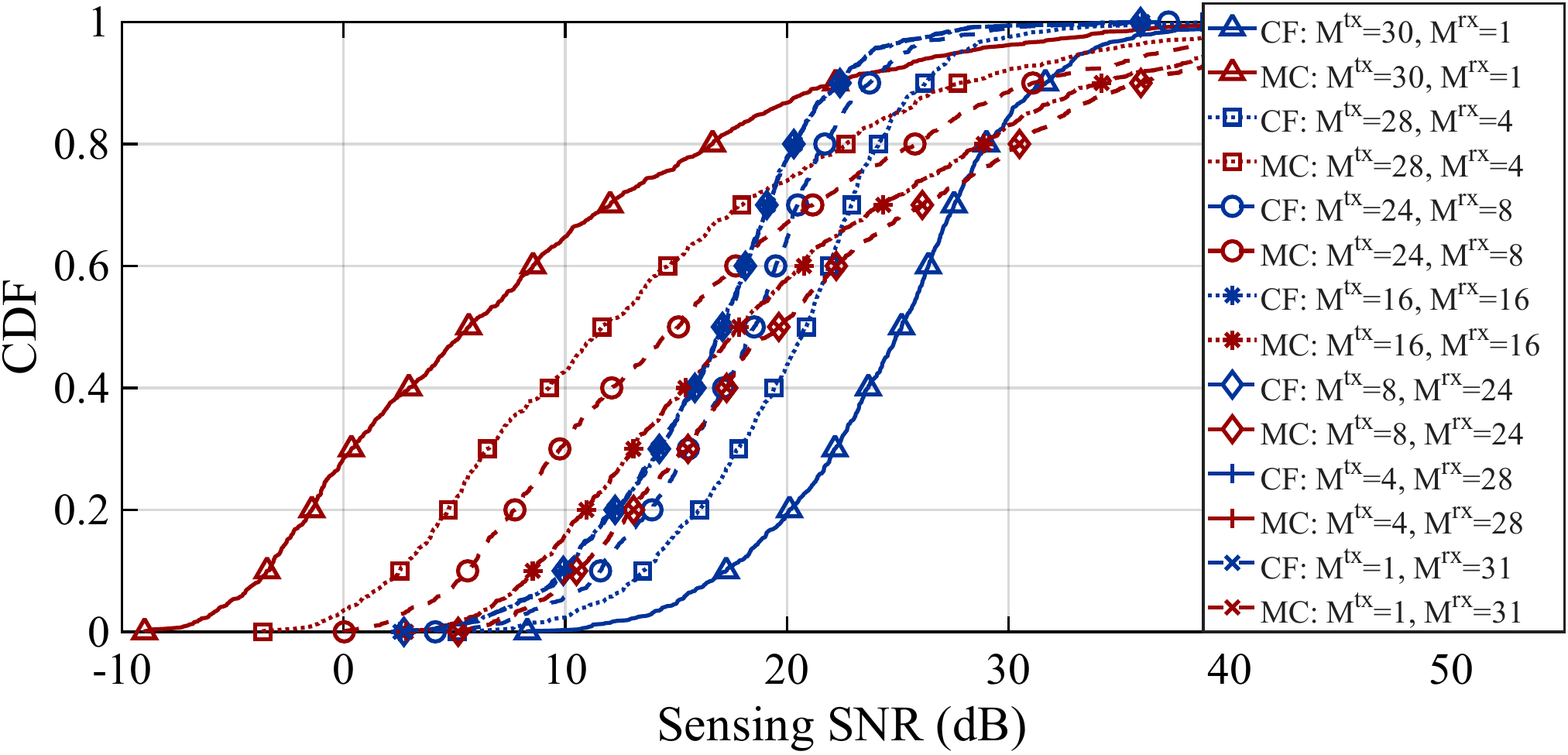}}
    \caption{Sensing SNR CDF for different transceiver allocation in both ISAC network frameworks.}
    \label{fig:4}
\end{figure}

To investigate how the distribution of transmit and receive resources affects sensing, we fixed $M^{\rm CF}=32$, $N_a^{\rm CF}=4$, and varied the Tx/Rx split ($\lvert{\calM^{\rm tx}}\rvert$ and $\lvert{\calM^{\rm rx}}\rvert$) in the CF system. This will accordingly vary the number of co-located Tx/Rx antennas in MC architecture, since $N_a^{\rm MC-tx}=\lvert {\calM^{\rm tx}}\rvert$, $N_a^{\rm MC-rx}=\lvert{\calM^{\rm rx}}\rvert$. 
The results depicted in Fig.~\ref{fig:4} reveal a clear structural trend in different extremes of allocation. For the maximum number of tAPs ($\lvert {\calM^{\rm tx}}\rvert=31$ and $\lvert {\calM^{\rm rx}}\rvert=1$), CF significantly outperforms MC, with a sensing SNR advantage of approximately $15$ dB. This is because CF benefits from distributed transmit diversity and favorable macro‑diversity, which become dominant when many APs contribute coherent energy toward the target. As the number of tAPs decreases and more APs are assigned to reception (e.g., $\lvert {\calM^{\rm tx}}\rvert=24$ and $\lvert {\calM^{\rm rx}}\rvert=8$ ), the performance gap narrows, yet CF outperforms MC in roughly $70\%$ of the cases. In the extreme case where transmission is highly concentrated ($\lvert {\calM^{\rm tx}}\rvert=1$ and $\lvert {\calM^{\rm rx}}\rvert=31$ ), MC becomes superior due to its large co‑located receive aperture and array gain; in this regime MC outperforms CF by about $14$ dB in $90\%$ of the cases.
This study highlights that CF excels when transmit resources are widely distributed, whereas MC benefits from concentrated receive apertures. The Tx/Rx allocation, therefore, plays a critical role in determining which architecture is more advantageous for sensing in network-level ISAC design. It is worth noting that \cite{buzzi2024scalability} reports that Tx/Rx split variations have negligible impact on CF communication performance.

\subsection{Distributed vs. Co‑Located: Antenna Deployment}  
\begin{figure}[t]
    \centerline{\includegraphics[scale = 0.27]{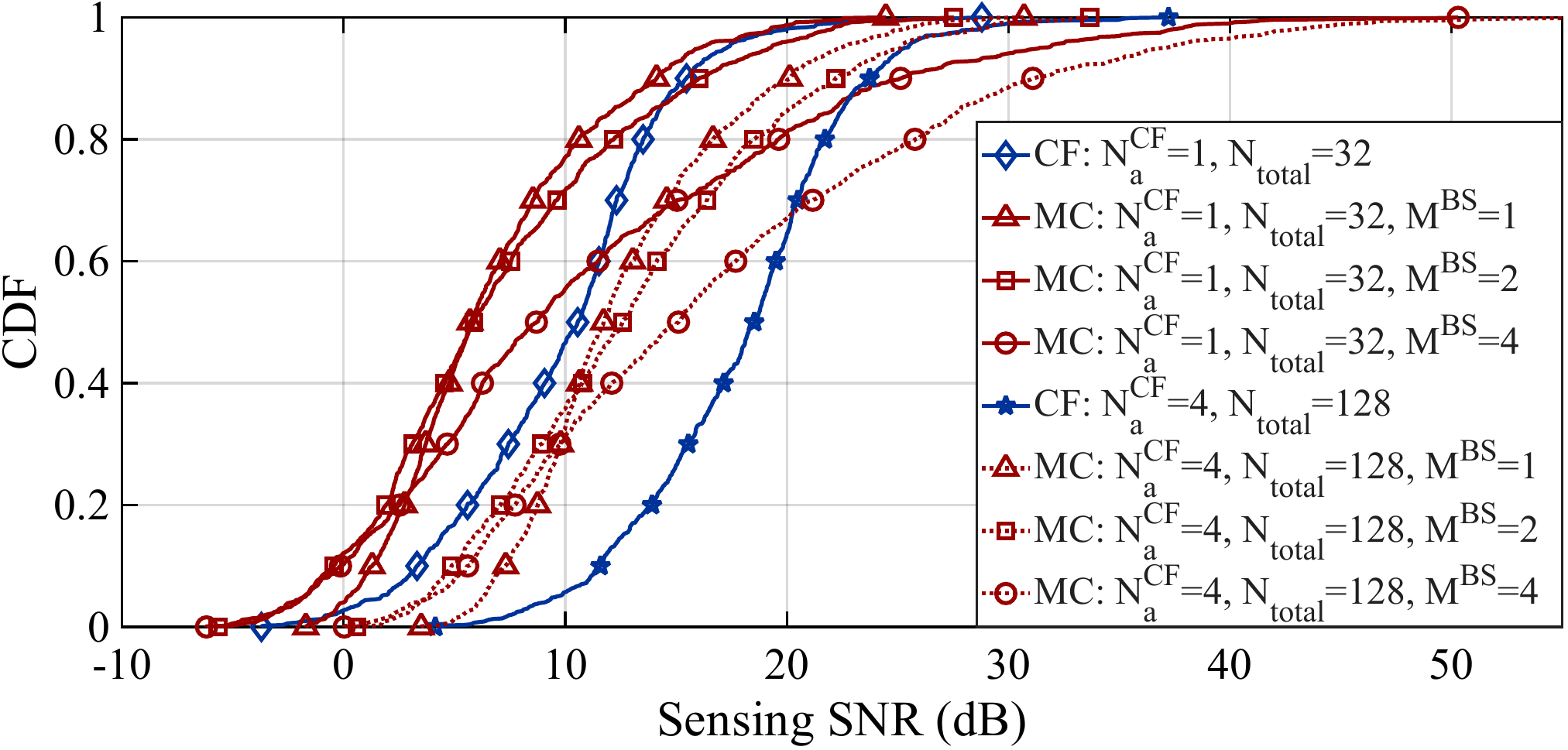}}
    \caption{Sensing SNR CDF comparison for CF and MC under $N_a^{\rm CF} =\{1,4\}$ and different number of BSs.}
    \label{fig:5}
\end{figure}

In this case study, we compare CF and MC sensing SNR under a fixed total number of antennas denoted by $N_{\rm Total}=M^{\rm CF}\times N_a^{\rm CF}$. Both architectures allocate $75\%$ of their antennas/APs to the transmission task, but the way these antennas are spatially arranged is fundamentally different. CF distributes its antennas across many APs, with $N_a^{\rm CF}=\{1,4\}$ elements, so each AP contributes only a modest amount of coherent sensing gain. In contrast, the MC architecture concentrates the same total number of antennas into a smaller number of BSs ($M^{\rm MC}=\{1,2,4\}$), with the per‑BS antenna count scaling as $N_a^{\rm MC} = N_{\rm Total}/M^{\rm MC}$. As $M^{\rm MC}$ increases from 1 to 4, BSs become more widely separated, effectively expanding the monostatic sensing aperture while preserving strong coherent beamforming at each site. The CDF analysis shown in Fig.~\ref{fig:5} reveals that CF outperforms MC with $4$ BS in a majority of realizations (about 60–70\% with different $N_a^{\rm MC}$ counts), highlighting that CF provides more robust sensing across random geometries. 

\section{Conclusions} \label{sec:7}
This work investigated the sensing performance of CF and MC OFDM-based ISAC architectures under a unified GLRT‑based detection framework. By deriving the corresponding test statistics and sensing SNR expressions for both systems, we highlighted the fundamental structural differences between distributed and co‑located transceiver deployments. Our numerical results consistently showed that CF‑ISAC benefits from the spatial diversity and geometric richness of distributed APs, leading to more stable and higher sensing SNR across a wide range of operating conditions.

\bibliographystyle{IEEEtran}
\bibliography{references}

\end{document}